\documentclass[preprintnumbers,prd,twocolumn,showpacs,floatfix,preprintnumbers,letterpaper,amsmath,amssymb,nofootinbib,superscriptaddress]{revtex4}
\usepackage{graphicx}
\usepackage{epsfig}
\usepackage{bm}
\usepackage{amsfonts}

\begin{document}

\title{ Non-minimally coupled $f(R)$ Cosmology}
\author{Shruti Thakur}
\email{shruti.thkr@gmail.com}
\affiliation{Department of Physics and Astrophysics, University of Delhi,
Delhi 110007, India.}

\author{Anjan A Sen}
\email{anjan.ctp@jmi.ac.in, asen@ictp.it}
\affiliation{Centre for Theoretical Physics, Jamia Millia Islamia, Delhi, India}
\affiliation{Abdus Salam International Center For Theoretical Physics, Trieste, Italy}

\author{T R Seshadri}
\email{trs@physics.du.ac.in}
\affiliation{Department of Physics and Astrophysics, University of Delhi,
Delhi 110007, India.}

\date{\today}
\begin{abstract}
 We investigate the consequences of non-minimal gravitational coupling
to matter and study how it differs from the case of minimal coupling
by choosing certain simple forms for the nature of coupling,
The values of the parameters are specified at $z=0$ (present epoch) and the equations are evolved backwards to calculate the evolution
of cosmological parameters. We find that the Hubble
parameter evolves more slowly in non-minimal coupling case as compared to the
minimal coupling case. In both the cases, the universe accelerates around present time, and enters the decelerating regime in the past.
Using the latest Union2 dataset for supernova Type Ia observations as well as the data for baryon acoustic oscillation (BAO) from SDSS observations, 
we constraint the parameters of Linder exponential model in the
two different approaches. We find that there is a upper bound on model
parameter in minimal coupling. But for non-minimal
coupling case, there is range of allowed values for the model parameter.
\end{abstract}
\maketitle
\section{Introduction}
\vspace{2mm}
It has now become fairly well established that the Universe is undergoing an accelerated expansion in recent times ($z \le 4$). 
Observational evidence for this mainly comes from supernovae Ia \cite{sn1a} and Cosmic Microwave Background Anisotropies \cite{cmb}. Large Scale Structure formation \cite{lss}, Baryon Oscillations \cite{bao} and Weak Lensing \cite{weak} also suggest such an accelerated expansion of the Universe.
One of the most challenging problems of modern cosmology is to identify the cause of this late time acceleration.
Many theoretical approaches have been employed to explain the phenomenon of late time cosmic acceleration.
A positive cosmological constant can lead to accelerated expansion of the universe but it is plagued by the fine tuning problem \cite{lcdm}. The cosmological constant may either be interpreted geometrically as modifying the left hand side of Einstein's equation or as a kinematic term on the right hand side with the equation of state parameter $w=-1$.
The second approach can further be generalized by considering a source term with an equation of state parameter $w<-1/3$. Such kinds of source terms have collectively come to be known as Dark Energy. Various scalar field models of dark energy have been considered in literature \cite{scalar1,scalar2,scalar3,scalar4,scalar5,scalar6,scalar7,scalar8,scalar9,scalar10,scalar11,scalar12,scalar13,scalar14,scalar15,scalar16,scalar17}. As an alternative to dark energy as a source for the accelerated expansion, modification of the gravity part of the action has also been attempted \cite{fr}. In these models, in addition to the scalar curvature $R$ in the gravity lagrangian there is an additional term $f(R)$. The gravity action hence becomes,
\begin{equation}
{\cal S} = {1\over{2 \kappa^2}} \int d^{4}x \sqrt{-g} [R + f(R)],
\end{equation}
However, in such models matter and gravity are still minimally coupled.
Despite the significant literature on such $f(R)$ models \cite{fr}, another interesting possibility which has not received due attention until recent times is a non -minimum coupling between the scalar curvature and the matter lagrangian density \cite{nonmin}.\\
In this paper we study the evolution of Hubble parameter in minimal and non-minimal coupling between scalar curvature and the matter lagrangian density. We use a form for $f(R)$ model which assumes the following exponential form proposed by Linder \cite{linder}:
\begin{equation}
 f(R)=-C\left[ 1-\exp(-R/R_0)\right] \label{lind_R}
\end{equation}
 with $C$ being the model parameter and $R_{0}$ is the present day curvature scale. We also attempt to place observational constraints on the parameters of this model in both minimal and non-minimal coupling of scalar curvature with matter lagrangian density. We find that there is an upper bound on model parameter $C$ in the minimal coupling case. For non-minimal coupling between matter and gravity, there is a range of values which the parameter $C$ is allowed to take. These bounds depend on the present day value of $q_0$ that we use.
The paper is organised as follows: In section \ref{FR} we introduce the most general action for modified gravity. The equations of motions corresponding to this action are solved numerically for both minimal and non-minimal coupling of scalar curvature with matter.  We investigate the observational constraints on the model parameter in section \ref{obscon}. In section \ref{conclusions} we summarize the results.
\section{$F(R)$ gravity models}   \label{FR}
\vspace{2mm}
We start with the general action for modified gravity where the
curvature is in general coupled with the matter lagrangian:

\begin{equation}
S = \int d^{4}x \sqrt{-g} \left[{1\over{2}} f_{1}(R) +
  [1+f_{2}(R)]{\cal{L}}_{m}\right],
\label{action}
\end{equation}
where $f_{1}(R)$ and $f_{2}(R)$ are the arbitrary functions of  Ricci
scalar $R$ and ${\cal{L}}_{m}$ is the lagrangian density for matter
which we will assume to be non-relativistic. We assume the
natural unit with the speed of light taken to be  unity in our calculations. The standard
Einstein-Hilbert action is recovered with $f_{2}(R)=0$ and $f_{1}(R) =
{R\over{\kappa^{2}}}$ where $\kappa^{2} = 8\pi G$.  The standard $f(R)$ gravity class of models are recovered with $f_{2}(R) = 0$ and $f_{1}(R)=
R+f(R)$, where $f(R)$ is an arbitrary function of $R$. In the latter case the pure gravity action has a non-minimal coupling while the matter is still minimally coupled to gravity. In what follows,
we shall consider the cosmological evolutions and their observational
constraints for modified gravity models in both cases, namely, one in which the curvature is coupled
minimally as well as the one in which the curvature is non-minimally coupled with the matter lagrangian
density.
\begin{figure*}[t]
\begin{tabular}{c@{\qquad}c}
\epsfig{file=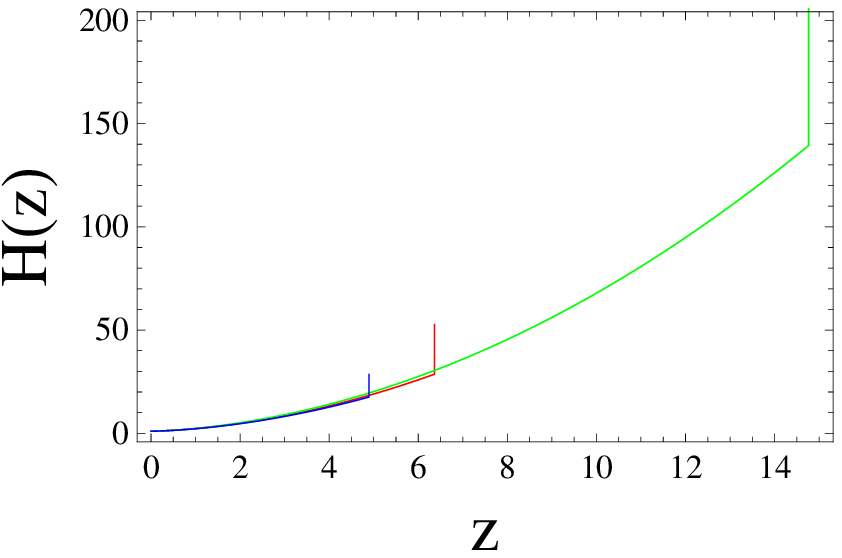,width = 7.5cm}&
\epsfig{file=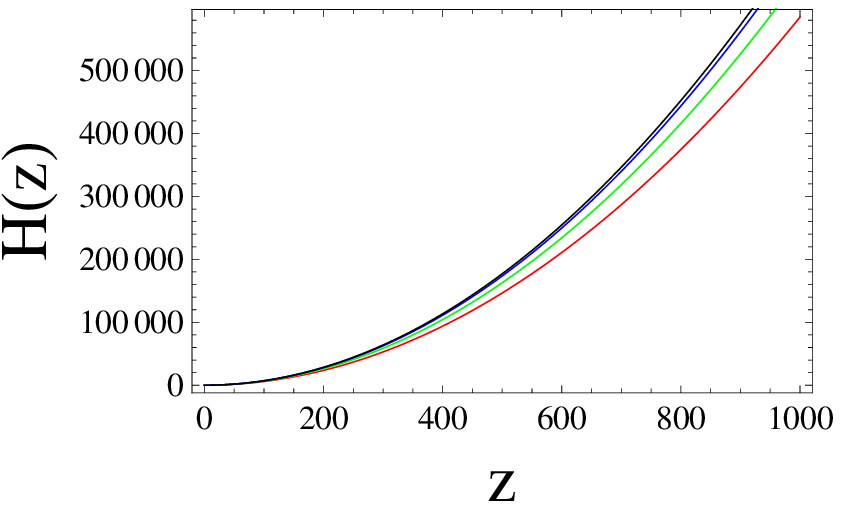,width = 8cm}
\end{tabular}
\caption{Behaviour of Hubble parameter as a function of redshift. $\Omega_{m}=0.25$ and $q_{0} = -0.55$ for both figures. In the left panel, $c = 1.1,1.2,1.4$ from left to right whereas in the right panel $c= 1.6,3,10,50$ from  right to left.}
\label{db_w_bf}
\end{figure*}
\subsection{Minimally Coupled f(R) gravity models}
\vspace{2mm}
As mentioned earlier, we recover the standard $f(R)$ gravity (i.e. the one with a minimal coupling of matter with gravity) if we have $f_{2}(R) = 0$ and $f_{1} = R + f(R)$ in the
action defined in equation (\ref{action}). Now varying the action given in equation (\ref{action})  with
respect to the metric tensor $g_{\mu\nu}$, we get the modified
Einstein equation:

\begin{equation}
G_{\mu\nu}+f_RR_{\mu\nu}-(\frac{f}{2}-\Box
f_R)g_{\mu\nu}-{\bigtriangledown}_\mu{\bigtriangledown}_\nu
f_R=\kappa^2 T_{\mu\nu}, 
\label{mineins}
\end{equation}
\vspace{2mm}
\noindent
where $f_R \equiv \frac{df}{dR}$ and $f_{RR} \equiv
\frac{d^2f}{dR^2}$. Assuming a flat Friedmann-Robertson-Walker
spacetime with a scale factor $a(t)$:

\begin{equation}
ds^2 = -dt^2 +a(t)(dx^2 + dy^2 + dz^2),
\label{frw}
\end{equation} 
\vspace{2mm}
the 0-0 component of the modified Einstein equation
(\ref{mineins}) becomes

\begin{equation}
H^2+\frac {f}{6} -f_{R}(HH^\prime +H^2)+H^2f_{RR}R^\prime
=\frac{\Omega_mH_0^2}{a^3},
\label{hmin}
\end{equation}
\vspace{2mm}
\noindent
where $\prime$ is with respect to $\ln a$, $\Omega_{m}$ is the matter
energy density parameter today and $H_{0}$ is the Hubble parameter
today. It will be convenient if we express the cosmological quantities involved in dimensionless units. Hence, we define the following dimensionless quantities:

\begin{eqnarray}
h &=& \frac{H}{H_{0}} \label{def}\\
x &=& \frac{R}{R_0} \\
{{\bar f}(x)} &=& \frac{f(R)}{R_{0}} \\
{{\bar f}_{x}} &=& \frac{d{\bar f}}{dx} =  \frac{df}{dR} = f_{R} \\
{{\bar f}_{xx}} &=& \frac{d^2 {\bar f}}{dx^2} = R_{0}\frac{d^2 f}{dR^2}=R_{0} f_{RR} \\
\alpha = \frac{R_{0}}{H_{0}^2} &=& 6(1-q_{0}),
\end{eqnarray} 
\vspace{2mm}
\noindent
where $R_{0}$ is the present curvature scalar and $q_{0}$ is the
present day deceleration parameter. The evolution of $R$ with redshift, $z$ is given by the equation,

\begin{equation}
\frac{dR}{dz}=6(3H\frac{dH}{dz}-(1+z)(\frac{dH}{dz})^2-(1+z)H\frac{d^2H}{dz^2}).
\end{equation}
\vspace{2mm}
\noindent
With these definitions, one can write equation (\ref{hmin})  in terms
of the redshift $z$, as,

\begin{eqnarray}
h^2+(1-q_{\circ}){f}+{f}_x((1+z)h\frac{dh}{dz}-h^2)+& &\nonumber\\
\frac{(1+z)^2}{1-q_{\circ}}h^3 {f}_{xx}(\frac{1}{h}(\frac{dh}{dz})^2
+\frac{d^2h}{dz^2} -\frac{3}{1+z}\frac{dh}{dz})
&=&{\Omega_{m}}(1+z)^3.\nonumber \\
\label{min}
\end{eqnarray}

Instead of working with a general $f(R)$, it will be more useful if we use a specific form for $f(R)$. A simple form is case of "Exponential Gravity"
proposed by Linder \cite{linder}. In this case the form of $f(R)$ is given by,
\begin{equation}
 f(R)=-C\left[ 1-\exp(-R/R_0)\right] \label{lind_R} 
\end{equation}
This can be expressed in the dimensionless form by defining a dimensionless constant $c=C/R_0$. We then have 

\begin{equation}
{\bar f}(x) = -c \left[ 1- \exp(-x)\right],
\label{lind}
\end{equation}
With this choice of ${\bar f}(x)$, we now solve
equation (\ref{min}) to find $H(z)$. There are a number of
investigations where this equation has been solved and the model is
subsequently constrained by observational data. In all of these works,
it is assumed that the universe behaves as a $\Lambda$CDM model in the
past and subsequently deviates from that behaviour \cite{amna}. In this way, one
sets the initial conditions for $h(z)$ and ${d h(z)}/{dz}$,
assuming the model is close to $\Lambda$CDM in the past. One problem
one usually faces in this approach, is the epoch at which to set the
initial conditions. Depending upon the redshift at
which one fixes the initial  conditions, one may or may not get
well-behaved solutions without any pathology. In other words, one has
to fine tune the initial conditions so as to get regular
solutions. Although this always happens in most of the relevant
modified gravity models discussed in the literature, it has not been
sufficiently emphasized to the best of our knowledge. To circumvent this problem, we take a
different approach to solve the equation (\ref{min}). We set the
initial conditions at present epoch, i.e at $z=0$. In equation
(\ref{min}), $h(z=0) = 1$ identically. Now for the second  initial 
condition, ${dh}/{dz} \mid_{z=0}$, one can write  ${dh}/{dz} (z=0)
= 1+q_{0}$, where $q_{0}$ is the present day deceleration parameter,
as mentioned before. Hence, the second initial condition is directly
dependent on the decelaration parameter at present and this will be
one of the parameters in our model. Hence, we have three parameters in
our model, i.e., $c$, which is the parameter in the Exponential
Gravity model, and two cosmological parameters, $\Omega_{m}$ and
$q_{0}$. Using these conditions for today's epoch, we evolve our system from present day (i.e., at
$z=0$) backwards in time (i.e., for increasing z). We want to stress that, in this approach, there is no extra assumption, or fine tuning of the initial conditions in order to solve the evolution equation (14).
One of the initial conditions, $h(z=0)$ is fixed to $1$ by definition and the other one is related to $q_{0}$, which we shall constraint by observational data. 

We aim to investigate if the cosmological evolution is well behaved as we go to earlier times, or whether they are plagued by singularities at high redshifts. We further aim to study the role of the parameter $c$ in this context.
As we discuss below, one indeed gets
regular solutions upto any higher redshift for a certain range of values for the
parameter $c$. For the rest of two parameters, we vary $\Omega_{m}$ between $0.25$ and $.0.35$ and the present deceleration parameter $q_{0}$ between $-0.9$ and $-0.55$.

We first investigate the behaviour of the normalized Hubble parameter ($h(z)=H(z)/H_0$) as function of redshift for $c < 1.6$  as shown in
figure 1. We have used three values for c, 1.1, 1.2 and 1.4 and we have assumed $\Omega_{m}= 0.25$ and 
$q_{0} = -0.55$. It clearly shows that singularity occurs at different redshifts for different values of $c$. Even if we vary $\Omega_{m}$ and $q_{0}$ in the range mentioned above, the overall behaviour does not change in the sense that there is always a singularity at some redshift. However, the behaviour significantly changes once we assume $c \ge 1.6$, as we show in figure 2. In figure 2 we have again plotted $h(z)$ as a function of $z$ but with the value of $c$ greater than 1.6. We evolve the system for redshift as high as $z=1000$, and the behaviour of $h(z)$ remains well-behaved. As we mentioned earlier, although these plots are for specific choices for $\Omega_{m}$ and $q_{0}$, the behaviour of $h(z)$ is similar for any value of these two parameters in the range mentioned above, i.e., $0.25 < \Omega_m < 0.35$ and $-0.9 < q_0 < -0.55$. So we can conclude that for the model parameter
$c > 1.6$, the model is regular upto any higher redshift.

We also study the behaviour of the deceleration parameter $q(z)$ as a function of redshift. In terms of the redshift, $q_0$ is given by,
\begin{equation}
q(z) = (1+z)\frac{H'(z)}{H(z0)}-1
\end{equation}
The behaviour of $q(z)$ is shown in figure 2 for different values of the model parameter $c$ assuming $\Omega_{m} = 0.25$ and $q_{0} = -0.55$. It shows that in all cases, the universe has an accelerated  phase at present and as we go back it smoothly enters the decelerating phase. The lower the value of $c$, the universe enters the accelerated  phase earlier.
\begin{figure}[t]
\epsfig{file=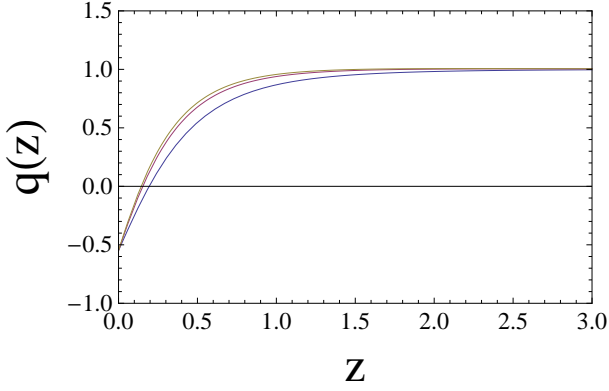,width=8 cm}
\caption{Behaviour of the deceleration parameter $q(z)$ as a function of redshift.  The lines are for $c = 2,5,10$ from bottom to top. we assume $\Omega_{m}=0.25$ and $q_{0}=-0.55$. }
\label{qmin}
\end{figure}

\vspace{5mm}

\subsection{Models with Non-Minimal matter-curvature coupling}

\begin{figure}[t]
\epsfig{file=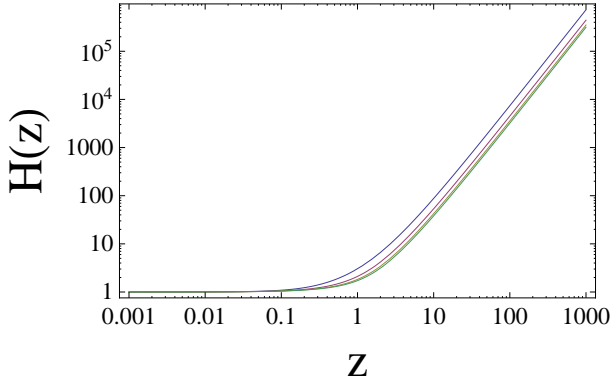,width=8 cm}
\caption{Behaviour of the Hubble parameter $q(z)$ as a function of redshift.  The lines are for $c = 1.5,6,15, 50$ from top to bottom. we assume $\Omega_{m}=0.25$ and $q_{0}=-0.55$. }
\label{qmin}
\end{figure}
Here  we extend the $f(R)$ gravity models assuming a non-minimal coupling between the matter lagrangian ${\cal L}_{m}$ and the scalar curvature. We assume the action to be,

\begin{equation}
{\cal S} = \int d^{4}x \sqrt{-g} \left[{1\over{2\kappa^{2}}} R + (1+f(R)){\cal L}_{m}\right]. 
\label{action2}
\end{equation}
\vspace{2mm}

Minimizing the above action with respect to $g_{\mu\nu}$ gives the modified version of the Einstein's Equation:

\begin{equation}
 \phi  R_{\mu\nu} -{1\over{2}}Rg_{\mu\nu} = \kappa^{2} (1+f(R)) T_{\mu\nu} + (\nabla_{\mu}\nabla_{\nu} - g_{\mu\nu}\Box)\phi,
\label{einnon}
\end{equation}
where $\phi = 1 + 2\kappa^{2}{{\cal L}_{m}}{df\over{dR}}$.  The Bianchi identity $G_{\mu\nu;\nu} = 0$ gives

\begin{figure}[t!]
\epsfig{file=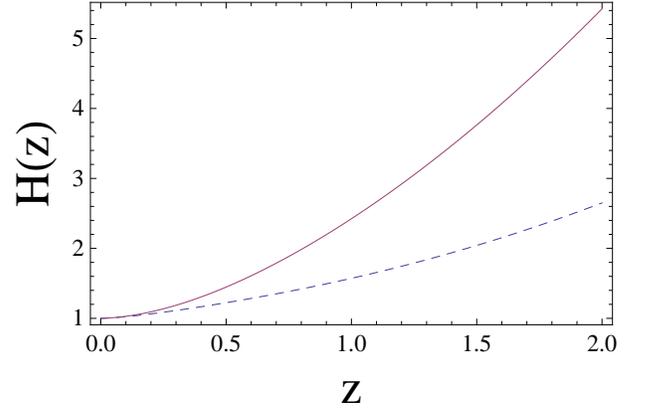,width=8 cm}
\caption{Behaviour of the Hubble parameter $q(z)$ as a function of redshift for minimally coupled (solid line) and the non-minimally coupled case (dashed line).  We assume $\Omega_{m}=0.25$ and $q_{0}=-0.55, c=5$. }
\label{qmin}
\end{figure}

\begin{equation}
 \nabla_{\mu}T_{\mu\nu} = {{df / {dR}}\over{1+f(R)}} (g_{\mu\nu} {\cal L}_{m} -T_{\mu\nu}) \nabla^{\mu}R,
\label{cons}
\end{equation}
which implies the non-conservation of the matter energy momentum tensor. This is due to the non-minimal coupling between the matter lagrangian and the curvature which results  exchange of energy between the matter and the scalar degrees of freedom present due to the $f(R)$ gravity model. This exchange of energy is one important feature of this non-minimally couple $f(R)$ gravity model. But  once we assume the prefect fluid form for our matter energy  momentum tensor (which is consistent with a homogeneous and isotropic universe) together with the form for the lagrangian density ${\cal L}_{m} = -\rho_{m}$ ($\rho_{m}$ being the matter energy density) \cite{rhom}, one can explicitly show that putting  $\nu=0$ in the above equation, i.e for the energy density conservation equation, one gets the usual equation, ${\dot\rho_{m}} + 3H\rho_{m} = 0$. With this together with the metric given by (\ref{frw}), one can write the 0-0 component of the equation (\ref{einnon}):
\begin{equation}
H \dot{\phi}+\frac{1}{6}f_1-(\dot H+H^2)\phi=
H_{\circ}^2f_2 \Omega_m(1+z)^3+H_{\circ}^2 \Omega_m (1+z)^3,
\label{hnon}
\end{equation}
\vspace{2mm}
where dot represents differentiation with respect to time. Further, comparing equation \ref{action} and \ref{action2}, $f_1$ is identified with $R/{\kappa^2}$ and $f_2$ with $f(R)$ given in equation \ref{lind_R}. Changing the variable to redshift $z$, and expressing everything in terms of the dimensional variables defined in (\ref{def}), one can now get from equation (\ref{hnon}) as

\begin{eqnarray}
((1+z)h\frac{dh}{dz}-h^2)(1-\frac{\Omega_m(1+z)^3
{\bar f}_{x}}{1-q_{\circ}})+(1-q_{\circ})x+\nonumber\\
h^3\frac{\Omega_m(1+z)^5{\bar f}_{xx}}{(1-q_{\circ})^{2}}
(\frac{3}{1+z}\frac{dh}{dz}-\frac{1}{h}(\frac{dh}{dz})^2-\frac{d^2h}{dz^2})\nonumber\\
+\frac{3(1+z)^3\Omega_m {\bar f}_{x} h^2}{(1-q_{\circ})} = \Omega_m(1+z)^3(1+{\bar f})
\label{nonmin}
\end{eqnarray}\vspace{2mm}
\begin{figure}[t]
\epsfig{file=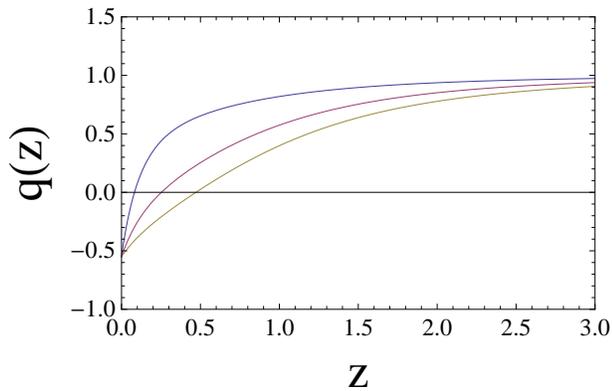,width=8 cm}
\caption{Behaviour of the deceleration parameter $q(z)$ as a function of redshift.  The lines are for $c = 2,5,10$ from top to bottom. we assume $\Omega_{m}=0.25$ and $q_{0}=-0.55$. }
\label{qmin}
\end{figure}

\begin{figure*}[t]
\begin{tabular}{c@{\qquad}c}
\epsfig{file=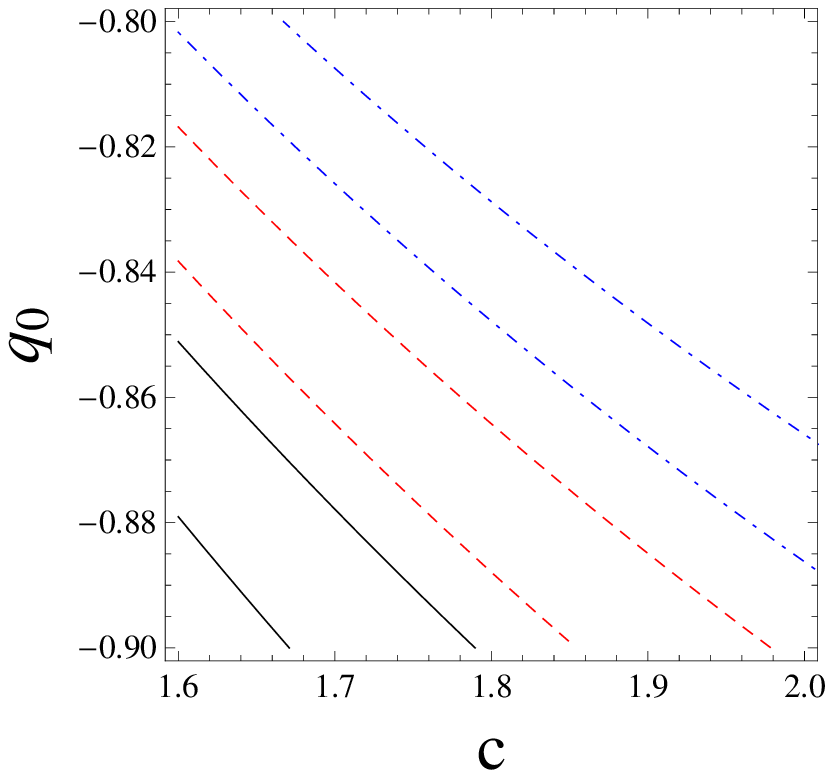,width = 8cm}&
\epsfig{file=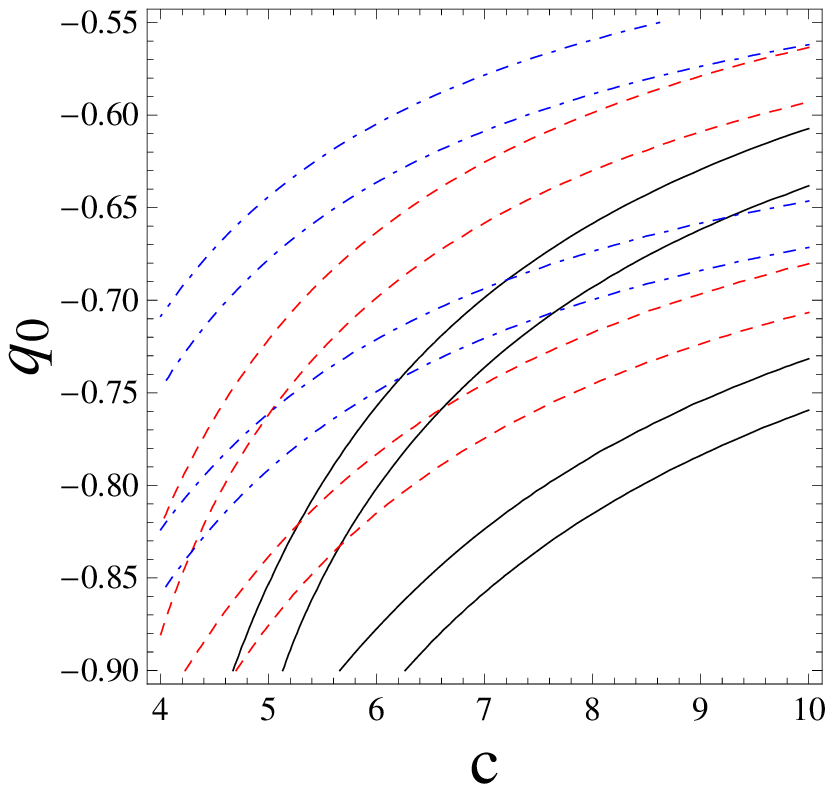,width = 8cm}
\end{tabular}
\caption{$1\sigma$ and $2\sigma$ contours in the $q_{0}-c$ plane for the minimally coupled  (left panel) and non-minimally coupled (right panel) case using the observational data (explained in the text). The solid, dashed and dash-dotted lines are for $\Omega_{m} = 0.25, 0.3, 0.35$ respectively.}
\end{figure*}
where subscript ``$x$'' denotes differentiation with respect to variable $x$.  The initial conditions are also taken in a way similar to the one described in the minimally coupled case in the previous section. The evolution of the Hubble parameter is shown in figure 3. Unlike the minimally coupled case, we do not have pathological behaviour for any values of the parameter $c$. To compare the minimally coupled and non-minimally couple cases, we show the behaviour of the Hubble parameter for the two case for same values of $c, \Omega_{m}$ and $q_{0}$ in figure 4. It shows that Hubble parameter in the non-minimally coupled case evolves slower than its counterpart in the minimally coupled case. 
We also show the behaviour of the deceleration parameter for nonminimally coupled case in figure 5. Here also the universe is in an accelerating phase at present and smoothly joins the decelerating regime in the past. Here unlike the minimally coupled case, higher the value of the parameter $c$, the universe enters the acceleration regime earlier.

\section{Observational Constraint}  \label{obscon}
In this section we investigate the observational constraints on  our model parameters. We use the Supernova Type Ia data from the latest Union2 dataset consisting of 557 data point \cite{union}. The data consists of the distance modulus defined as ${\cal \mu}=m-M=5{\rm log}d_{L}+25$ where  $d_{L}(a)$ is the luminosity distance defined as,
\begin{equation}
d_{L}(a)={a}^{-1} \int_{a}^{1}\frac{{\rm d}y}{y^{2}H(y)}.
\end{equation}
The other data we consider is the baryon acoustic oscillations scale produced in the decoupling surface by the interplay between the pressure of the baryon-photon fluid and gravity. For this we calculate the distance ratio $D_{v}(z=0.35)/D_{v}(z=0.2)$ where $D_{v}$ is given by

\begin{equation}
D_{v}(z) = \left[{z\over{H(z)}}\left(\int^{z}_{0}{dz'\over{H(z')}}\right)^2\right]^{1/3}.
\end{equation}

The SDSS observation gives $D_{v}(z=0.35)/D_{v}(z=0.2) = 1.736 \pm 0.065$ \cite{ratio}. We use this measurements together with the measurements of distance modulus by Type Ia supernova observations, to constrain our model.

We use these two observational data to put constraints on our model parameter. In Figure 6, we show the $1\sigma$ and $2\sigma$ contours in the  $q_{0}-c$ plane with different choices of the density paremeter $\Omega_{m}$.  For minimally coupled case, there is always an upper bound for the parameter $c$ as well as for the present day decceleration parameter $q_{0}$. For example, with $\Omega_{m}= 0.25$ the  $2\sigma$ upper bound for $c$ is around 1.8. This upper bound shifts towards higher values as one increases $\Omega_{m}$. 
On the other hand, for non-minimally coupled case, things are different. Here there is an allowed range of $c$ for every values of $q_{0}$. For example, with $\Omega_{m} = 0.25$ and $q_{0}= -0.9$, the  $2\sigma$ allowed range for $c$ is between $4.6$ and $6.2$.  But as one increases $\Omega_{m}$, this range shifts towards smaller value of $c$.   Also for the minimally coupled case, our constraint on parameter $c$ differs than that obtained by Ali et.al \cite{amna} They solved the evolution equation in a different way than ours. They have assumed the universe is close to $\Lambda$CDM at higher redshifts, and put the initial conditions in the early time whereas we put the initial condition at present and solve it backwards. Also one of the initial conditions $q_{0}$ is actually one of the fitting parameters. 
Also they have used the slightly older Supernova data given by constitution set, which has 397 data points in comparison to our 557 data points given by the Union2 set. Their analysis shows no constraint on $c$. at any level. It differs significantly from what we obtain.
\vspace{5mm}
\section{Conclusion} \label{conclusions}
We have reinvestigated the $f(R)$ gravity model where the curvature is minimally as well as in the case where gravity is non-minimally coupled with matter. We have assumed the Linder's Exponential form for $f(R)$ for our analysis. We have fixed the initial condition at present. We do not need any additional assumptions. By definition,  $H/H_{0} = 1$ at $z=0$, where $H_{0}$ is the present day Hubble parameter. The other initial condition fixes $q_{0}$, the present day deceleration parameter, and we take $q_{0}$ as one of our fitting parameters. Hence our method of solving the evolution equation does not involve any additional assumption.
First we check that for minimally coupled case, for $c \ge 1.6$, one can evolve the universe till any higher redshifts without any pathological behaviour. For lower values of $c$, there is some singular features in $H(z)$ at different redshifts depending upon the parameter choices. For nonminimally coupled case, this changes and $H(z)$  is regular till any higher redshifts for any choice of parameter values (we have taken $\Omega_{m}$ between 0.25 and 0.35, $q_{0}$ between -0.9 to -0.55 and $c$ between 1 to 50). For minimally coupled case, the constraint $c \ge 1.6$ to get regular solutions for $H(z)$ also satisfies the constraint coming from the local gravity tests \cite{fr}. Also the evolution of the universe is as expected, showing accelerating universe at late time, and decelerating universe in the past.
We next use the the observational data coming from Type Ia supernova observations as well as the BAO peak measurements by SDSS. For supernova, we use the latest Union2 compilation consisting 557 data points. For minimally coupled case, the constraints on $c$ is completely different from what obtained by Ali et al. \cite{amna} earlier. We have obtained a upper bound on $c$. This upper bound on $c$ shifts towards the higher values as one increases $\Omega_{m}$. For nonminimally coupled case, there is a range of allowed values for $c$, and this allowed range for $c$ shifts towards for smaller  values as one increases $\Omega_{m}$.

\section{Acknowledgement}
A.A.S acknoweledges the financial support privided 
by the University Grants Commission, Govt. Of In-
dia, through major research project grant (Grant No:33-
28/2007(SR)). A.A.S also acknowledges the financial support provided by the Abdus Salam International Center For Theoretical Physics, Trieste, Italy where part of the work has been done. ST and TRS acknowledge facilities provided by Inter University Center For Astronomy and Astrophysics, Pune, India through IUCAA Resource Center(IRC) at Department of Physics and Astronomy, University of Delhi, New Delhi, India.

\end{document}